# INTERACTIVE METHODS OF TEACHING PHYSICS AT TECHNICAL UNIVERSITIES

Ľuboš Krišťák, Miroslav Němec, Zuzana Danihelová

*Abstract*: the paper presents results of "non-traditional" teaching of the basic course of Physics in the first year of study at the Technical University in Zvolen, specifically teaching via interactive method based on an increased focus on problem tasks and experiments. This paper presents also research results of the use of the given method in conditions of Slovak universities and its comparison with traditional methods.

## 1 Introduction

In the 1980's David Hestenes and Ibrahim Halloun (Halloun, I., Hestenes, D., 1985) from the USA published papers on didactic research, whose object were students of secondary schools and universities, dealing with misconceptions in Newtonian dynamics. One of the research results was a test (Force Concept Inventory (FCI)) (Hestenes, D. et al., 1992) containing questions from Newtonian mechanics connected to everyday life. The authors decided to research whether the students understand the basic concepts from mechanics sufficiently; how they are able to work with them and apply them into various everyday situations. The test results from the whole world showed (Slovak version Hanč et al., 2008) that the traditional teaching of the Newtonian mechanics in the early years of university study eliminates wrong perception of students, acquired during their secondary school studies, only to a small extent. It was also shown that traditional lectures or seminars help to acquire only basic knowledge without deeper understanding and to algorithm solving of problems; the students do not show conceptual understanding of the subject which should result from sufficient number of solved quantitative tasks and from logically clear lectures (Redish, E.F., 2003, Hanč, J. et al., 2008). Next important conclusion of using these tests was that the misconceptions (not only in mechanics) and their accumulation in further study cause that students do not understand the subject dealt with and that they are learning the subject by heart what consequently leads to frustration.

## 2 Innovative Methods in Teaching of Physics

In last three decades various interactive methods have become very popular. Their use brings about much better results than the use of traditional methods. One of these methods is modern approach which was developed at the Institute of Physics at the University of Dortmund. The essentials of this approach are that a better education of physics teachers must put more emphasis on: the teaching of educational philosophy as well as individual preconceptions in the minds of pupils, avoiding and overcoming misconceptions, the deliberate use of mental processes such as assimilation and accommodation, the cognitive conflict as a trigger for changes of thought structures, more simple and qualitative experiments done by learners, exercises to improve comprehension, the making explicit of the connection between formalism and the real world, and the recognition of the role of the affective domain in the physics teaching-learning process. These elements concern several components of the teaching-learning process: didactic principles and educational findings, pedagogical strategies and understanding of subject matter, department and interdisciplinary orientation, teacher´s self-concept and student´s motivation, intellectual growth and emotional development. All these components are interconnected and their integration leads to a better education for future physics teachers (Nachtigall, D.K., 1990).

Some of other these methods are PI (Peer Instruction), ILD (Interactive Lecture Demonstration), JiTT method (Just-in-time-teaching), etc. (Mazur, E., 1997, Crouch, C.H., Mazur E., 2001). These methods emerge mainly from the interactivity between the lecturer and students, whereas students are actively involved into individual stages of the teaching and learning process and actively participate in solving of the dealt problems what gives an immediate feedback to the lecturer and he/she can immediately respond to incorrectly understood concepts, or misconceptions (e.g. Sokoloff, D.R., Thornton, R.K., 1997).

The meaning of the word "to know" has changed from "be able to remember" to "be able to find information and use it" (Simon, H.A., 2006, Stebila, 2010). Research into the area of Physics methodology among other things has shown that an increased focus on experimenting during the teaching and learning process and the use of qualitative (problem) tasks encourages students to solve problems and look for new procedures in discovering information (Hockicko, 2010, Holbrook, 2009, Žáčok, 2010). The use of creative experiments in the teaching process increases the level of understanding and attention of students and at the same time the theory of physics is becoming interconnected with everyday life (Bussei, 2003, Dykstra, 1992, Zelenický, 1999). The use of qualitative tasks from Physics supports the fixation of knowledge and at the same time these tasks enable to test the knowledge and practical skills. Such tasks influence also increased interests of students in the subject and support active understanding and application of curriculum within the teaching process. They are very precious when developing physical thinking (Bednařik, Lepil, 1995). While solving a qualitative task students must dive into the issue or phenomenon. In the process they often realise that they do not understand the phenomenon as well as they thought they did (misconceptions). A great advantage of qualitative tasks is the practical application of theoretical knowledge. While solving qualitative tasks students learn to analyse the phenomena, develop logical thinking, sense and creativity (Němec, 2008).

## 3 Interactive Method Based on Increased Focus on Problem Tasks and Experiments (Interactive P&E Method)

We tried to build on the benefits of using multimedia, experiments and qualitative tasks in the teaching of Physics. The result is the interactive P&E method whose main idea is interactive working with students with the help of experiment and problem tasks analyses. It is able to use this method during lectures as well as seminars. In the case of lectures, before the lecture the lecturer sets basic terms or concepts which have to be explained and prepares several problem tasks for each area. After explaining of a particular concept a problem task follows (or several problem tasks connected to the given concept). This problem task is given to students by the teacher; this way the interactivity is provided. A discussion with analysis of all possible solutions follows (teacher is a moderator). If the task is more difficult the lecturer can use experiment, simulation, applet, video-experiment or video-analysis (e.g. in the programme Tracker that is free available) (Hockicko, 2011) to steer the students in the correct answer. A similar situation is in the case of demonstration experiment – the lecturer carries out an experiment which is then connected to a problem task. A great advantage of this method is that students are engaged in all stages of the lecture and that they are forced to thing and concentrate. At the same time they learn how to argue and analyse individual problem situations; teacher has an immediate feedback and has the possibility of correcting students' misconceptions resulting from their incorrect answers, or students among themselves can correct these misconceptions.

An example of using this method in hydrostatics: after explaining Archimedes' principle a problem tasks follows on the lecture. Students supervised by the lecturer discuss this task and come to adequately reasoned conclusion. During a test such a task can be eventually used as a multiple choice.

How does a ship's draught change after shipping out from a river into the sea?

a) it increases, b) it decreases, c) it remains the same, d) from given data it is not possible to give a clear answer.

In a similar way also theoretical seminar form Physics is conducted; besides calculating of traditional (quantitative) tasks the teacher integrates also problem or combined tasks (partially calculation and partially problem tasks) into the process. The analysis is similar as in the case of lectures.

Combined task: Mass of a lift cage with passengers is 500 kg. How will the lift move if the force of the rope is 5 kN? Do not consider friction force. ($g = 10 m.s^{-2}$) Task can be supplemented by questions about the direction of acceleration and instantaneous speed, eventually provide multiple choices on the test:

a) lift will not move as the force of the rope and gravity of the lift with passengers is the same,

b) lift fill move upwards with accelerated motion,

c) lift will move downwards with accelerated motion,

d) lift will be at rest or in uniform motion, however from the instruction it cannot be said in which direction.

**4 Research**

**Research object**

Students of the first year of study at the Faculty of Environmental and Manufacturing Technology (FEVT) were the object of research. Interactive teaching method was used within the basic course of Physics. This method was based on increased focus on problem tasks and experiments followed with an analysis (P&E) (Krišťák, Němec, 2011).

Pedagogical experiment was carried out at the Technical University in Zvolen during the academic year 2009/10 at the Faculty of Environmental and Manufacturing Technology. Experiment was applied to the subject Physics in the first year of Bachelor degree of study with two lectures and two seminars per week. Considerable part of the contents of Physics in the first year of study at the University is aimed at revision and deepening of knowledge acquired during secondary school. This knowledge is adequately extended by higher Physics that should be managed by the students of the first year at university. The primary goal of the subject Physics in the first year of study at the Technical University in Zvolen is to minimize the differences in level of students' knowledge acquired at secondary school (Danihelová, A., 2006).

Four control groups and four experimental groups took part in the experiment at the Faculty of Environmental and Manufacturing Technology in the winter semester. In the control groups the teaching process was performed in traditional way. The traditional way of teaching means thirteen lectures and thirteen seminars within a semester. Thirteen seminars are aimed at calculating of exercises (quantitative tasks) from individual areas of Physics, which were dealt within the lectures.

In experimental groups the students took part in thirteen lectures and thirteen seminars, too, whereby the interactive P&E method was used on both, lectures and seminars.

**Research Objectives**

The main research objective was the comparison of educational results reached in the teaching process using interactive P&E method and results reached in the traditional way of teaching. Knowledge at the four levels of learning (remembering, understanding, specific – use knowledge in typical situations and nonspecific transfer – use knowledge in problem situations) was researched.

To reach the aim partial tasks were determined:
- to verify stated hypotheses using research tools and methods;
- to find out whether method based on problem tasks and experiments influences the level of knowledge of students in the first year of study at the Faculty of Environmental and Manufacturing Technology in the subject Physics.

**Research Hypotheses**

Based on the aforementioned goal the main hypothesis was formulated:
H: The use of interactive P&E method in the teaching of Physics in the first year of study influences the level of student's knowledge from Physics significantly.

To verify the main hypothesis operational hypotheses were stated:

$H_1$: At the end of the experimental teaching process students taught by the interactive P&E method achieve higher performance in didactic test in the area of specific transfer than students taught traditionally.

$H_2$: At the end of the experimental teaching process students taught by the interactive P&E method achieve higher performance in didactic test in the area of nonspecific transfer than students taught traditionally.

$H_3$: At the end of the experimental teaching process students taught by the interactive P&E method achieve higher performance in didactic test in the area of remembering than students taught traditionally.

$H_4$: At the end of the experimental teaching process students taught by the interactive P&E method achieve higher performance in didactic test in the area of understanding than students taught traditionally.

**Research Methods and Techniques**

To achieve the stated objectives and to verify hypotheses following research methods and techniques of empirical research were proposed:

- pedagogical experiment

- didactic test (DT) for verifying operational hypotheses $H_1$-$H_4$ (see appendix)

- statistical methods for research results processing.

**Selection of Respondents**

The research was carried out in four experimental and four control groups. These groups contained 140 students taking part in the course of Physics; students were divided into the groups randomly. For the research purpose the results of all control groups were joined into one control group and the results of all experimental groups into one experimental group, both with 70 students.

After the selection of suitable groups the pedagogical experiment followed simultaneously during one year in all groups. During the experiment students did not know that they were a part of an experiment. After the experiment pupils in all groups took didactic test. All students had the same didactic test (there were only two groups that had different task order) containing 30 questions. Knowledge at the four levels of learning (remembering, understanding, specific and nonspecific transfer – use knowledge in typical and problem situations) was researched (Table 1). Tasks in the test were in accordance with the curriculum of Physics for high schools; such curriculum corresponds to the contents of the subject Physics. Task structure was the same as in the tests of ŠPÚ (National Institute for Education of Slovak Republic) or CERMAT (Centrum for Evaluation of Educational of Czech Republic). All questions were multiple choices with four offered solutions.

After the pedagogical experiment obtained data was collected and statistically and qualitatively analysed.

Tab.1 Four levels of learning in used test

|  | Level of learning | | | |
| --- | --- | --- | --- | --- |
|  | Remembering | Understanding | Specific transfer | Nonspecific transfer |
| Points | 1 | 2 | 3 | 4 |
| Question in test | 1,3,5,7,8,10,13,14,15,23,28 | 4,9,11,19,21,24,30 | 6,17,18,26, 29 | 2,12,16,20,22,25,27 |

**Research Results**

To verify hypotheses $H_1$-$H_4$ a non-standardised didactic test - posttest, taken by students at the end of the semester, was used (see appendix).

Normal distribution was verified via Kolmogorov-Smirnov Test. Results in the chart (Fig. 1) show that there is a difference between knowledge of students in experimental and control group showed in the test. Statistical verification of hypothesis was carried out using two-sample T-test and F-test. At first F-test was used to assess the equality of variances. Following, the hypothesis of equally acquired score in the control and experimental groups was tested. Independent two-sample Student T-test for large populations and equal variances was used.

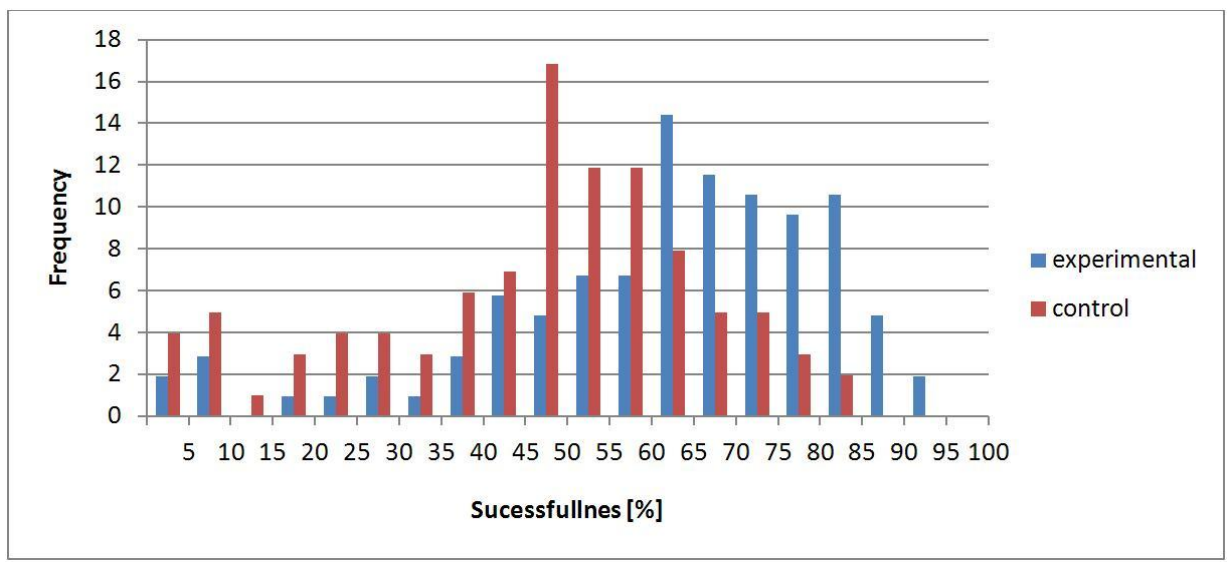

**Figure 1 Test successfulness histogram in the control and experimental group. (control group: N=70, Mean = 44.41%, Stand. Dev. = 22.73%, Max = 82%, Min = 2%, experimental group: N =70, Mean = 57.72%, Stand. Dev. 22.03%, Max = 94%, Min = 1%).**

**Summary of Research Results**

Validity of operational hypotheses is summarized in Tab. 2.

**Tab. 2 Summary of individual hypotheses verification**

| Hypothesis | Method of data acquisition | Hypothesis validity | Researched value |
|:---:|:---:|:---:|:---:|
| $H_1$ | DT | **valid** | Specific transfer |
| $H_2$ | DT | **valid** | Nonspecific transfer |
| $H_3$ | DT | **valid** | Remembering |
| $H_4$ | DT | **valid** | Understanding |

From statistical analyses and results of partial hypotheses testing it is possible to say that the initial hypothesis is confirmed and true. Research into the use of P&E method in the teaching process shows that students taught by the interactive P&E method achieved higher performance in the didactic test in the area of specific and nonspecific transfer, understanding and remembering at the end of the experimental education than students taught traditionally.

**5 Conclusion**

Implementation of qualitative tasks in the teaching of Physics contributes to the application of basic didactic principles: e.g. it increases the principle of demonstration; students are forced to participate in the teaching process more actively (principle of activity). Students (even those who will not tackle Physics in detail) can use skills and experience acquired during solving of qualitative tasks also in further study at the university. Observations also imply that students are more attentive and active during qualitative tasks closely connected to practice and everyday life. Also students who are not interested in Physics very much and achieve worse results participate in solving such tasks. In a similar way, the use of experiments in the teaching process improves demonstration of the curriculum, increases students' attention, forces them to work and think independently and helps to show the connection between physical theory and everyday life in natural technique and society.

The testing of students confirmed that if we want to achieve better results with current student quality, it is inevitable to replace traditional methods with new, interactive methods which are commonly used at foreign universities with the technical focus.

We used our experience with the teaching via P&E method also while creating two textbooks from Physics for the students of the first year of study at technical universities (Gajtanska et al., 2012, Bahýl et al. 2013).

**This paper was elaborated with the support of the projects KEGA no. 005UMB-4/2011 and KEGA no. 011UMB-4/2012.**

**PaedDr. Ľuboš Krišťák, PhD.[1]**
**Mgr. Miroslav Němec, PhD.[1]**
**Mgr. Zuzana Danihelová[2]**
**Department of Physics, Electrical Engineering and Applied Mechanics[1]**
**Faculty of Wood Sciences and Technology, Technical University in Zvolen**
**Institute of Foreign Languages, Technical University in Zvolen[2]**
**T.G. Masaryka 24**
**960 53 Zvolen**
**Tel: +521455206836**
**E-mail: kristak@tuzvo.sk**
      mnemec@acoustics.sk
      zuzana.danihelova@tuzvo.sk

Appendix 1: DIDACTIC TEST

1 Vector physical quantity is: A) matter, B) time, c) momentum, D) mean velocity.

2 The graph in the picture describes the train's motion before entering the station. What was the brake acceleration? **A)** $\frac{40}{12} m.s^{-2}$, **B)** $\frac{30}{12} m.s^{-2}$, **C)** $\frac{30}{10} m.s^{-2}$, **D)** $\frac{30}{8} m.s^{-2}$

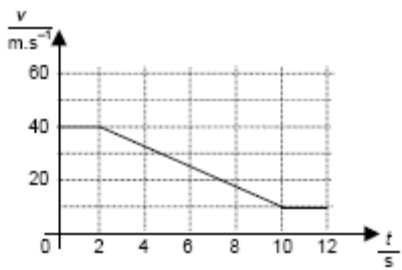

3 Dimensions of Unit of force are: A) kg.m.s, B) kg.m.s$^{-1}$, C) kg.m.s$^{-2}$, D) kg.m$^2$.s.

4 Friction force <u>does not depend</u> on: A) roughness of the surfaces, B) amount of the surface area, C) normal force, D) coefficient of friction.

5 Dimensions of Joule are: A) kg.m$^2$.s$^{-2}$, B) kg.m.s$^{-2}$, C) kg.m$^2$.s$^{-3}$, D) kg.m$^2$.s$^{-3}$.

6 If the force has a vertical direction on the shift direction, work will be determined by the equation: a) W = 0, b) W = F.s, c) W = F.t, d) W = -F.s.

7 Gravity is between: A) electrically charged bodies, B) celestial bodies, C) bodies of huge size, D) physical bodies

8 If an object has an initial horizontal velocity, moves on a part of: A) straight line, B) circle, C) parabola, D) ellipse.

9 For the centre of mass it is not true that: A) each body has exactly one, B) it cannot be outside the body, C) it is a physical centre of the body, D) it is affected by final gravitational force.

10 Kinetic energy of a body in a rotary movement is defined as:

A) $E_k = \frac{1}{2} J.\omega^2$, B) $E_k = \frac{1}{2} m.v^2$, C) $E_k = \frac{1}{2} m.r^2$, D) $E_k = m.g.h$

11 The principle of mass conservation for fluid flowing is expressed by: A) continuity equation, B) Bernoulli equation, C) Stoke's law, D) Pacsal's law.

12 Compare the magnitudes of lift affecting copper and lead bodies with the same volume when submerged into water: A) copper body is affected by a greater lift, B) lead body is affected by a greater lift, C) both lifts are the same, D) it cannot be stated.

13 Internal energy is: A) Sum of kinetic energy and potential energy of the body, B) Sum of kinetic energy and potential energy of all body parts, C) Product of kinetic energy and potential energy of the body, D) Product of kinetic energy and potential energy of all body parts.

14 The unit of heat capacity is: A) $J.kg^{-1}.K^{-1}$, B) $J.kg^{-1}$, C) $J.K^{-1}$, D) J

15 Dimensions of molar gas constant are: a) $J.kg^{-1}.K^{-1}$, b) $J.K^{-1}.mol^{-1}$, c) $J.K^{-1}$, d) $J.mol^{-1}$.

16 In two tanks there are molecules of hydrogen and chlorine at the same temperature. Which of the molecules have lower root-mean-square-speed? A) chlorine, B) hydrogen, C) both are the same, C) it cannot be stated.

17 Bicycle frame is deformed mainly by: A) strain, B) tension, C) shear, D) bend.

18 For the coefficient of thermal expansion of iron and concrete it is true that: A) coefficient of thermal expansion of iron is higher, B) coefficient of thermal expansion of iron is lower, C) both coefficients are comparable, D) both coefficients are the same.

19 What is not true for surface tension? A) its unit is $N.m^{-1}$ B) it depends on the matter, C) it increases with increased temperature, D) it does not depend on the surface energy.

20 Mercury in a glass in weightlessness: A) fills the whole glass also from outside, B) creates a spherical shape, C) spills across the bottom, D) remains in original state.

21 Saturated steam: A) has a higher temperature than gas, B) is in equilibrium with its liquid, C) is created by isothermal increase of the volume of superheated steam without the presence of liquid, D) has always the same density as its liquid.

22 How is called a part of phase diagram where solid matter, its liquid and their saturated steam coexist? A) triple point, B) critical point, C) saturated steam curve, D) superheated steam curve.

23 Which statement is not true for electric charge? A) it can be transported within a body, B) it is a physical quantity, C) it is always bound to an atom, D) it is positive or negative.

24 During parallel connection of resistors: A) their total resistance is higher than the resistance of any of them, B) resistor with lower resistance has higher heat energy, C) electrical current through each of them is equal regardless of their resistance, D) higher voltage is on the resistor with higher resistance.

25 Thinning of the light bulb filament results in: A) lowering of input power due to the lowering of filament resistance, B) lowering of input power due to the increase of filament resistance, C) increase of input power due to the lowering of filament resistance, D) increase of input power due to the increase of filament resistance.

26 Total energy of an oscillator is: A) constant, B) equal to the sum of kinetic and potential energy, C) equal to the remainder of kinetic and potential energy, D) equal to the product of kinetic and potential energy.

27 What will be the frequency of and oscillator if its mass will decrease 9 times? Its initial frequency was 81 Hz. A) 9Hz, B) 27Hz, C) 243 Hz, D) 729 Hz.

28 Wave length is the distance between: A) the nearest points oscillating in the same phase, B) neighbouring nods, C) the nearest amplitudes, D) neighbouring antinode and node.

29 Quantities describing oscillation are the function of: A) place only, B) time only, C) place and time, D) neither place, nor time.

30 In the following time diagram a record of a tone recorded from two microphones M1 and M2 is illustrated. Which statement is not true?

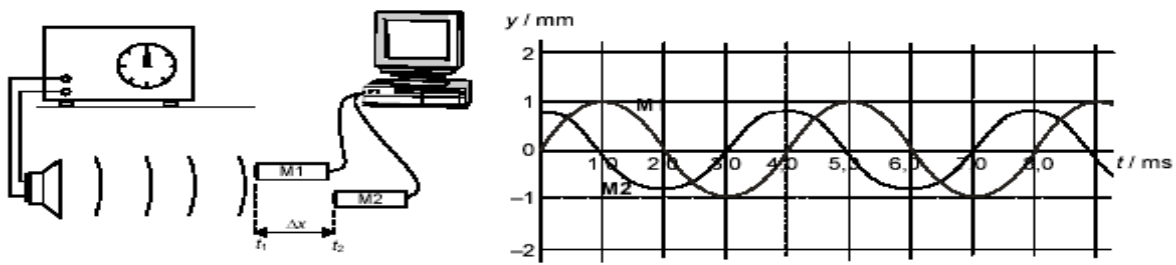

A) Membrane of the M2 microphone oscillates with smaller amplitude than M1,
B) Membrane of the M2 microphone is later in the phase than M1 membrane by $\frac{\pi}{2}$,
C) Period of the tone recorded by microphones is 4ms,
D) Height of both tones is 250Hz.